# Optimal sizing of solar photovoltaic and lithium battery storage to reduce grid electricity reliance in buildings

Han Kun Ren
Department of Engineering Science
University of Oxford
7 Keble Road
OX1 3QG Oxford
United Kingdom
han.ren@eng.ox.ac.uk

Malcolm McCulloch
Department of Engineering Science
University of Oxford
Begbroke Science Park, Sandy Lane
OX5 1PF Yarton
United Kingdom
malcolm.mcculloch@eng.ox.ac.uk

David Wallom
Department of Engineering Science, University of Oxford
7 Keble Road
OX1 3QG Oxford
United Kingdom
david.wallom@oerc.ox.ac.uk

## Keywords

renewable energy, optimisation, hybrid, photovoltaics, electric storage

## Abstract

In alignment with the Paris Agreement, the city of Oxford in the UK aims to become carbon neutral by 2040. Renewable energy help achieve this target by reducing the reliance on carbon-intensive grid electricity. This research seeks to optimally size solar photovoltaic and lithium battery storage systems, reducing Oxford's grid electricity reliance in buildings. The analysis starts with modeling the electricity demand. The model uses Elexon electricity settlement profiles, and assembles them into the demand profile according to the quantity and types of buildings in Oxford. Then, solar generation is modeled using Pfenninger and Staffell's method. Solar photovoltaic and lithium storage systems are sized using a hybridized analytical and iterative method. First, the method calculates the solar system size search range, then iterates through the range. At each solar size, the method calculates and iterates through the storage system size search range. Within each iteration, the renewable system is simulated using demand and generation data with a simplified system set-

## Nomenclature

| Symbol | Description | Symbol | Description |
|---|---|---|---|
| $C$ | critical point matrix | $n_i$ | number of companies with class $i$ demand |
| $C_{PV}$ | solar PV system capacity (MW) | $\eta_c$ | storage charge efficiency |
| $C_{PVMax}$ | max solar PV system capacity (MW) | $\eta_d$ | storage discharge efficiency |
| $C_s$ | storage system capacity (MWh) | $n_{PV}$ | solar PV system lifespan (year) |
| $D$ | electricity demand (MW) | $n_S$ | storage system lifespan (year) |
| $D_i$ | class $i$ electricity demand (MW) | $N$ | total number of companies |
| $DoD$ | depth of discharge (%) | $O_{PV}$ | solar PV system O&M cost (EUR/MW/year) |
| $E$ | storage system size (MWh) | $O_S$ | storage system O&M cost (EUR/MWh/year) |
| $f_i$ | class $i$ correction factor | $P_E$ | electricity wholesale price (EUR/MWh) |
| $F$ | solar generation capacity factor (MW/MW) | $r$ | discount rate (%) |
| $G$ | solar generation (MW) | $S$ | storage level (MWh) |
| $I_{PV}$ | solar PV system installed cost (EUR/MW) | $S_{real}$ | real storage level (MWh) |
| $I_S$ | storage system installed cost (EUR/MWh) | $S_{sustain}$ | sustainable starting storage level (MWh) |
| $L$ | lower triangular matrix | $t$ | time (hour) |
| $LCOE$ | levelized cost of electricity (EUR/MWh) | $\Delta t$ | time step (hour) |
| $M$ | difference matrix | $t_n$ | time at the n$^{th}$ critical storage level (hour) |
| $m_i$ | number of electricity meters in Class $i$ | $T$ | time horizon (hour) |
| $M$ | total number of electricity meters | | |

up and the conventional operation strategy. The method outputs combinations of solar system capacity, storage system capacity, and grid electricity import. Each combination's levelized cost of electricity is calculated, and the lowest cost combination is the optimal sizing. Solar and storage system costs are projected from 2019 to 2100, and the optimal sizing is calculated for each year. The result shows that solar photovoltaic is economically competitive, but lithium storage cost is still too high. As solar and storage prices continue to drop, they will take up greater portions of the energy system. However, there will always be a need for the grid, as it provides flexibility and can meet demands that are too costly for solar and storage.

## Introduction

In alignment with the Paris Agreement, the city of Oxford in the UK aims to become carbon neutral by 2040 (Oxford City Council, 2021). Renewable energies are clean and environmentally friendly, suitable for achieving the carbon-neutral target. The renewable industry has seen tremendous growth, with solar photovoltaic costs reduced by 85 % since 2010 (IRENA, 2020). In many countries, these reduced costs enable solar PV (photovoltaic) farms to produce electricity significantly cheaper than fossil fuel power plants. Furthermore, solar is also more accessible, providing power for remote communities unreachable by the grid. The environmental, economic, and accessible merits give solar generation an edge over fossil fuel power plants. As a result, solar PV installations currently account for 98 % of Oxfordshire's renewable generation (Low Carbon Hub, 2021). A significant drawback of solar is its temporal and weather dependency, causing energy generation to be intermittent and non-dispatchable, unable to follow the varying demand. Energy storage can fix these issues by storing energy during surplus generation and releasing that energy during excess demand. Therefore, this paper aims to help the city of Oxford determine the optimal size of solar PV and lithium battery systems, reducing buildings' reliance on grid-supplied electricity.

The sizing of hybrid renewable systems has four main approaches: iterative, heuristic, mathematical optimization, and analytical (Anoune et al., 2018). The iterative approach systematically iterates through a range of renewable sizes, simulating the renewable system at each size, and selecting the optimal size based on system performance and cost. Ma used HOMER (Hybrid Optimization of Multiple Energy Resources) software to size solar PV panels, wind turbines, and batteries for a Hong Kong island with an average electricity demand of 250 kWh/day (Ma et al., 2014). The study found the lowest cost solar-wind-battery system consists of 145 kW solar, 21 kW wind, and 705 kWh battery capacity. Borowy proposed a hybrid iterative and analytical method to size solar-battery systems (Borowy & Salameh, 1994). The iterative method produces solar and battery size combinations that can meet a specific amount of electricity demand, measured in LPSP (Loss of Power Supply Probability). Then, the relationship between solar and battery sizes is modeled with an analytical equation, and the equation is solved for the lowest-cost solar and battery sizes. Cabral used an iterative method to size a solar-battery system (Cabral et al., 2010). The study used Markov chain and probabilistic beta distribution to create solar generation profiles. The profiles are fed into an iterative method, yielding solar and battery sizes based on the LPSP requirement, following which the lowest cost size is selected as optimal.

The heuristic approach uses algorithms, such as genetic and particle swarm algorithms, to navigate the size search space smartly, allowing faster convergence to near-optimal. Yang used a genetic algorithm to size a solar-wind-battery system for a telecommunication relay station in Guangdong (Yang et al., 2009). The method found that a 2 % LPSP system needs 7.8 kW solar, 18 kW wind, and 120 kWh battery capacity. The system was built, and the energy records showed the system met the 2 % LPSP requirement, demonstrating the method's validity. Shabani also used a genetic algorithm to size a solar-wind-storage system (Shabani et al., 2020). The study found complementarity between wind and solar generation can reduce the storage requirement. Maleki used the Monte Carlo method with particle swarm algorithm to size a solar-wind-battery system (Maleki et al., 2016). The study found the wind-battery system has the lowest cost due to lower storage requirements.

Mathematical optimization algorithms, such as linear programming, uses objective and constraint equations describing the renewable system to optimize sizing. Atia used mixed-integer linear programming to size a grid-connected solar-wind-battery system (Atia & Yamada, 2016). The study used Gaussian distribution to model solar generation and demand, and Weibull distribution to model wind generation. The results showed demand flexibility could reduce storage requirements. Lai used optimization algorithms to size solar farms with anaerobic digestion plants and energy storage (Lai & McCulloch, 2017). The study explored seven optimization algorithms, and found particle swarm optimization with the interior point method was the most suitable.

The analytical approach models relationships between renewable size and system behavior, and then sizes based on the desired behavior. Arun proposed an analytical method to size storage in a solar-diesel-battery system (Arun et al., 2010). The study first simulates the storage energy level profile. The storage profile is constrained to positive energy levels with the same start and end. Then, storage is sized according to the highest energy level in the profile. Bandyopadhyay proposed an improved version of Arun's method (Bandyopadhyay, 2011). First, the storage profile is shifted upwards to eliminate negative storage levels. Then, the latter section of the profile is shifted downward to equalize the start and end. Finally, the storage is sized according to the largest storage level.

Each of the four main approaches has its advantages and drawbacks. Heuristic algorithms can smartly navigate the search space to reduce calculation time, but they are stochastic and cannot guarantee the exact optimal. On the other hand, some mathematical optimization algorithms can find the exact optimal, but they require a complex set of constraint equations to describe the renewable system. Similarly, the analytical approach can also find the exact optimal, but analytical equations are difficult to formulate. The iterative approach is simpler, but it has a long calculation time due to the large search space, and it cannot find the optimal outside the search space. Hybridizing the approaches can mitigate their drawbacks and harness their advantages (Yang et al., 2018).

This paper presents a hybridized analytical and iterative method to size solar photovoltaic and lithium storage systems, reducing grid reliance in buildings. The method uses analytical equations to calculate the maximum solar and storage system sizes to define the size search space. The search space guarantees optimal while reducing the iterative method's calculation time. The contributions of this paper are summarized as follows:

- This paper utilizes novel analytical equations to calculate maximum solar and storage system sizes, which are used to define the size search space. The defined search space guarantees that optimal sizing is within it, and reduces the number of iterations for the iterative method.

- Most studies do not consider the future trend of solar PV and lithium battery composition in the energy system. This paper analyzes the future trend with solar and battery system cost projections.

- This paper presents a new method to model building electricity demand for districts in the United Kingdom.

The paper is structured as follows. The paper first introduces the hybrid renewable system, the component models, and the iterative sizing method. Then, the sizing results and discussions are presented. Finally, the last section concludes the paper.

## Methodology

### SYSTEM SETUP AND OPERATION STRATEGY

A hybrid renewable energy system is proposed to reduce Oxford's reliance on the grid. The renewable system is modeled using a simplified system setup with the conventional operation strategy, as shown in Figure 1. The simplified setup consists of a common AC (Alternating Current) bus regulated by the electricity grid. Demand directly taps into the AC bus. Solar PV generation needs to convert from DC (Direct Current) to AC via an inverter before connecting to the bus. Lithium battery storage operates in DC, and needs a bi-directional AC/DC converter to connect to the bus. The conventional operation strategy states that solar generation will first be used to meet the demand. Any excess generation will be stored in lithium battery storage. When generation cannot meet the demand, storage energy will be used. When both generation and storage are not enough, electricity will be imported from the grid to meet the demand. The availability of these three supply sources ensures the demand is met at all times.

### ELECTRICITY DEMAND

Oxford's building electricity demand is modeled using Elexon electricity settlement profiles. The profiles were constructed using half-hourly demand data sampled from Great Britain (Elexon, 2018). There are 120 Elexon profiles, divided into eight classes of domestic and commercial consumers (UKERC, 2021). Within each class, the profiles are divided into five seasons: winter, spring, summer, high summer, and autumn. And within each season, the profiles are divided into three types of days: Weekday, Saturday, and Sunday.

For each class, the Elexon profiles are compiled into an hourly demand profile spanning a year based on the day of the week and the season. Then, Oxford's hourly demand profile is compiled using Eq. (1), which says the district's hourly demand profile ($D(t)$) is the sum of each class' demand profile ($D_i(t)$) scaled by the number of meters in that class ($m_i$) and the correction factor ($f_i$). The BEIS (UK's Department of Business, Energy & Industrial Strategy) provides statistics on the number of meters in each district (BEIS, 2019). The number of domestic meters for Class 1 and 2 are provided individually. However, the number of non-domestic meters in Class 3 to 8 are lumped together. Since Class 3 to 8 model non-domestic consumers with increasing demand levels, the analysis assumes the demand level is related to the company size in terms of employees. For example, demands for small companies are modeled by Class 3, and large companies by Class 8. The ONS (UK's Office of National Statistics) provides statistics on the number of companies in each district and their sizes (ONS, 2019). The number of meters for Class 3 to 8 is determined using Eq. (2). The equation says the number of meters in a particular class ($m_i$) is equal to the total number of non-domestic meters ($M$) scaled by a ratio. The ratio is equal to the number of companies in that class ($n_i$) divided by the total number of companies ($N$).

$$D(t) = \sum_{i=1}^{8} D_i(t) m_i f_i \quad (1)$$

$$m_i = M \frac{n_i}{N}, where\ N = \sum_{i=1}^{6} n_i \quad (2)$$

Elexon profiles were originally developed in 1997 (Elexon, 2018). Since then, domestic demands have decreased due to more efficient appliances, while non-domestic demands have increased due to new technologies and increased business activities. To account for these changes, correction factors ($f_i$) are calculated using linear regression. First, demand profiles are assembled for all districts in the UK, and annual demands are calculated from these profiles. Then, the calculated demands and actual demands are fed through linear regression to obtain the correction factors, shown in Table 1. The normal distribution of company sizes in each district causes high collinearity in company size data. To work around the collinearity, annual demands from Class 3 to 8 are combined and fed through linear regression, yielding a single correction factor. The correction factors reduce the discrepancy between actual and calculated annual demands by more than 90 %, as suggested by the R-Square values. The P-Values are less than 0.05, suggesting significant correlations between calculated and actual demands. Lower and Upper 95 % values define the lower and upper bounds of the 95 % confidence interval where correction factors reside.

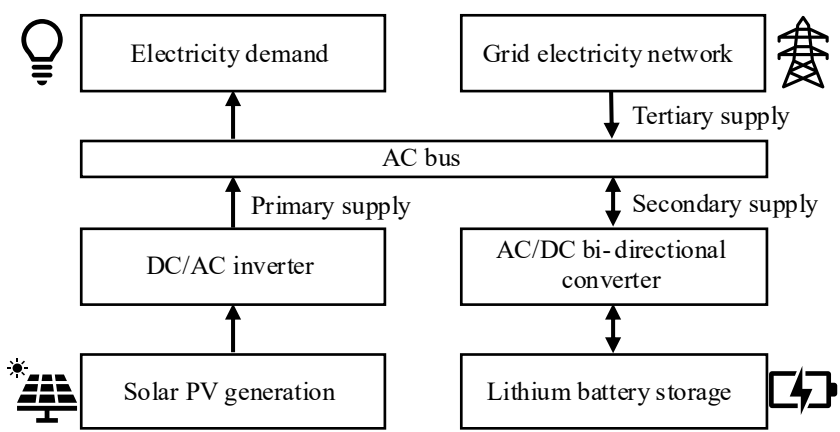

*Figure 1. Proposed hybrid renewable energy system setup in Oxford.*

Table 1. Correction factors from linear regression.

| Class | Correction Factor ($f_i$) | R-Square | P-Value | Lower 95% | Upper 95% |
|---|---|---|---|---|---|
| Profile Class 1 (Domestic) | 0.84 | 0.99 | 0 | 0.83 | 0.85 |
| Profile Class 2 (Domestic) | 0.72 | 0.96 | 0 | 0.70 | 0.73 |
| Profile Class 3-8 (Non-Domestic) | 2.65 | 0.90 | 0 | 2.56 | 2.73 |

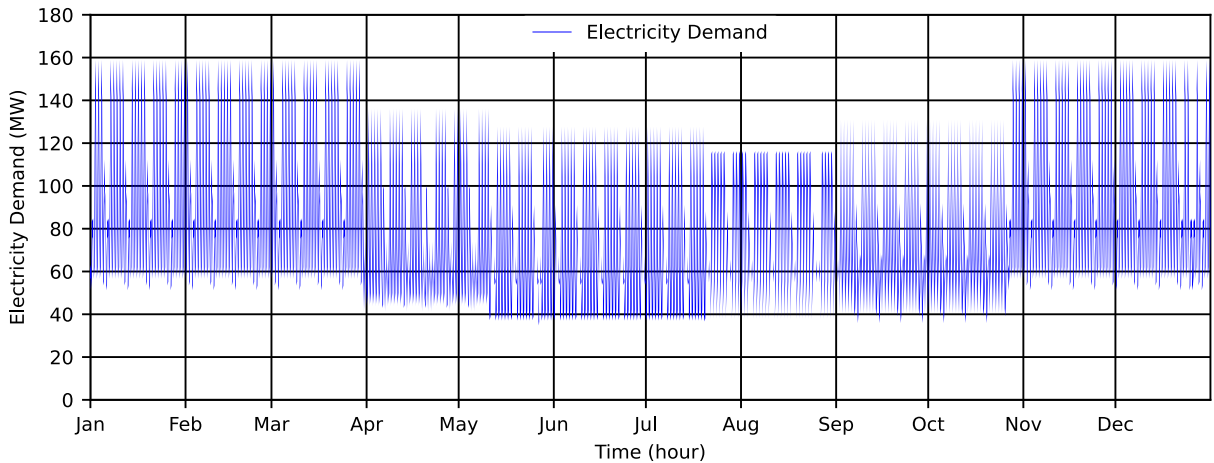

*Figure 2. Oxford's hourly electricity demand profile in a year.*

These correction factors are applied in Eq. (1) to form Oxford's demand profile in Figure 2. The demand profile shows slight peaks at night from Economy 7 meters. Then demand rises in the morning, peaks at noon, and drops in the afternoon. It peaks again at 5 pm when the working hour ends, then slumps in the evening. Demand is highest during weekdays and lowest on Sundays, because non-domestic demands dominate in Oxford. Furthermore, the demand is the highest during winter and lowest during high summer, and demand variation is also greater during winter. The present approach is benchmarked against another approach, where real-time national hourly demand data is scaled down according to the district's population. When compared, the present approach's annual demand is 9 % closer to the actual value than the benchmark approach.

**SOLAR GENERATION**

Solar capacity factor is the ratio of actual energy output to the rated energy output of the PV panel. It is calculated using Pfenninger and Staffell's method via the Renewables.ninja service (Pfenninger & Staffell, 2016). The simulation is set at Oxford, with a system efficiency loss at 10 %, mostly due to DC-to-AC inverter (Pfenninger & Staffell, 2016). The panel orientation is set at a 30 degrees tilt facing south, suitable for European countries (Jacobson & Jadhav, 2018). The simulation first obtains solar irradiance and temperature data from NASA's MERRA database. The temperature data is fed into Huld's temperature-dependent efficiency curve to get solar panel efficiency (Huld et al., 2010). The irradiance data is processed into direct and diffused irradiance using the Boland-Ridley-Lauret model (Ridley et al., 2010). Direct and diffused irradiance are then scaled according to the solar panel tilt. The capacity factors are calculated using system efficiency, panel efficiency, direct irradiance, and diffused irradiance. The hourly capacity factors are simulated from a five-year average (2015 to 2019) to reduce anomalies. Figure 3 shows the capacity factor profile. The capacity factors are only non-zero during sunlight hours, as solar panels only generate electricity in sunlight. Furthermore, the capacity factor peaks at noon due to high sunlight intensity, and is higher during summer due to higher direct irradiance.

Eq. (3) uses the capacity factor profile to create the solar generation profile. The equation says solar generation profile ($G(t)$) is equal to the capacity factor profile scaled by the solar PV system capacity ($C_{pv}$). The iterative method needs a maximum solar PV system capacity to set the solar size search range, and Eq. (4) calculates this. The equation says the maximum solar PV system capacity ($C_{PVMAX}$) is equal to the maximum ratio of demand ($D(t)$) divided by solar capacity factor ($F(t)$); given that the ratios are from all hours ($t$) of the year ($T$) where the capacity factor is greater than zero. In essence, the equation says for each hour with solar generation, find the solar PV capacity that can meet the demand during that hour, and then select the maximum solar PV capacity.

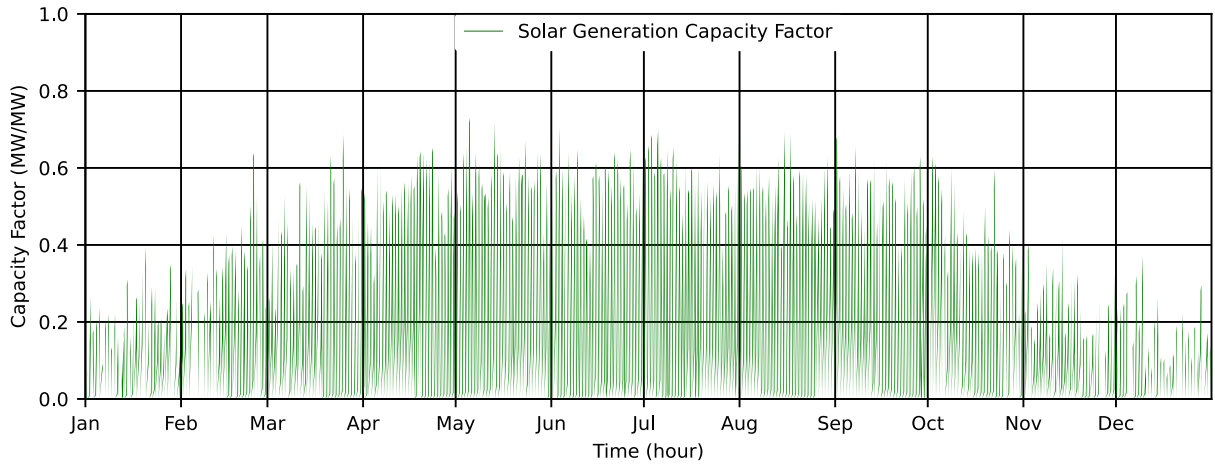

*Figure 3. Oxford's hourly solar generation capacity factor profile in a year.*

$$G(t) = F(t)C_{PV} \quad (3)$$

$$C_{PVMax} = Max\left(\frac{D(t)}{F(t)}\right) | t \in T, F(t) > 0 \quad (4)$$

### BATTERY STORAGE

The conventional operation strategy states when solar generation is greater than electricity demand, the surplus generation is charged into lithium battery storage; when generation is less than demand, the excess demand is met via energy discharged from storage. The resulting storage behavior is modeled by Eq. (5). The equation says storage level at a future time ($S(t+\Delta t)$) is equal to the current storage level ($S(t)$) plus the change in storage level. The change in storage level is equal to the difference between generation ($G(t)$) and demand ($D(t)$), multiplied by the storage efficiency ($\eta(t)$) and the time elapsed ($\Delta t$). The storage efficiency is the charge efficiency ($\eta_c$) during surplus generation ($G(t) - D(t) > 0$), as surplus generation is charged into storage. It is one over the discharge efficiency ($1/\eta_d$) during excess demand ($G(t) - D(t)$ 

< 0), as excess demand is met via energy discharged from storage. In this analysis, the storage system consists of lithium batteries and a bi-directional converter, with an assumed charge and discharge efficiency of 0.8 (Cabral et al., 2010).

A storage profile is constructed by starting at zero, and calculating the storage level at the end of each time step using Eq. (5). The size of the storage system is governed by the largest storage level difference in the profile, calculated using critical points. Eq. (6) calculates the critical points by setting the derivative equal to zero and finding the roots. Moreover, computer functions, such as "argrelextrema," find critical points by comparing each point with neighboring points. A critical point's storage level is higher or lower than neighboring points'.

The storage level differences used in storage sizing are captured by matrix $M$ in Eq. (7). In essence, matrix $M$ pairs critical points and calculates the storage level difference between them; $L+C$ produce the first half of the pairs, and $C^T$ produce the second half. Matrix $C$ is created via the dot product of the all-ones vector's transpose (($\mathbf{1}_n$)$^T$) and the critical storage levels vector ($C_n$). The lower-left half of matrix $L$ is filled with the difference between starting and ending storage levels ($S(T) - S(0)$). Matrix $L$ pushes the lower-left half of matrix $C$ into the second analysis period, as critical storage levels in the second period can also govern the sizing.

$$S(t + \Delta t) = S(t) + \left(\big(G(t) - D(t)\big)\eta(t)\right)\Delta t, where: \eta(t) = \begin{cases} \eta_c, if\ G(t) - D(t) > 0 \\ \dfrac{1}{\eta_d}, if\ G(t) - D(t) < 0 \end{cases} \quad (5)$$

Solve $\dfrac{dS(t)}{dt} = 0$ for all $t$, such that critical points are: $t_1, t_2, t_3, \dots, t_{n-1}, t_n$,
and critical storage levels are: $S(t_1), S(t_2), S(t_3), \dots, S(t_{n-1}), S(t_n)$ \quad (6)

$$M = L + C - C^T \quad (7)$$

$$Where: L = \begin{cases} l_{i,j} = 0, if\ i \leq j \\ l_{i,j} = S(T) - S(0), if\ i > j \end{cases}$$

$$C = (\mathbf{1}_n)^T C_n$$

$$C_n = [S(t_1), S(t_2), S(t_3), \dots, S(t_{n-1}), S(t_n)], and\ \mathbf{1}_n = [1,1,1,\dots,1]$$

Eq. (8) states that storage system size ($E$) depends on the difference matrix $M$ and the overall storage profile trend. If the overall storage profile is increasing, that is, the ending storage level ($S(T)$) is greater than the starting storage level ($S(0)$), then storage size is equal to the absolute value of the difference matrix's minimum ($|Min(M)|$), corresponding to the storage profile's largest decrease. Conversely, if the overall storage profile is decreasing ($S(T) - S(0)<0$), then storage size is equal to the difference matrix's maximum ($Max(M)$), corresponding to the stoage profile's largest increase. Lastly, if the overall storage profile is neither increasing nor decreasing ($S(T) - S(0)=0$), then storage size is equal to the maximum absolute value of the difference matrix ($Max|M|$), corresponding to the largest change in the storage profile.

$$E = \begin{cases} |Min(M)|, \ if \ S(T) - S(0) > 0 \\ Max(M) \ , \ if \ S(T) - S(0) < 0 \\ Max|M| \ , \ if \ S(T) - S(0) = 0 \end{cases} \quad (8)$$

The equation yields the optimal storage system size. When compared, undersized storage cannot keep up with the demand and empties sooner due to the smaller size. It also stays empty for extended periods, during which the storage cannot supply energy, forcing the system to import grid electricity. In contrast, oversized storage never empties, and as a result, part of the capacity is wasted and never used. In comparison, the optimally sized storage is the maximum storage size without any wasted capacity. Thus, the optimal storage size will set the maximum in the storage size search range.

With the storage sized, a realistic storage profile can be constructed using Eq. (5) while constrained by Eq. (9). Eq. (9) says real storage levels ($S(t)_{real}$) cannot exceed storage size ($E$) or fall below zero. Once the storage level hits the storage size, no more energy can be charged into storage until space becomes available. On the other hand, if storage is empty, energy cannot be extracted until it is charged again.

$$0 \leq S(t)_{real} \leq E \quad (9)$$

A real storage profile can repeat itself indefinitely when subject to the same demand and generation conditions in the future. Equal starting and ending storage levels characterize this repeatability, and they are calculated using Eq. (10). The equation says sustainable starting storage level ($S(0)_{sustain}$) is equal to the real storage profile's ending storage level ($S(T)_{real}$). Starting at the sustainable storage level enables storage to end at the same level, ensuring storage operation's future sustainability.

$$S(0)_{sustain} = S(T)_{real} \quad (10)$$

The lifespan of energy storage mainly depends on the amount of usage. Storage manufacturers specify the depth of discharge for lifespan-related design. For example, an 80 % depth of discharge with 6,000 cycles means the storage can be charged and discharged for 6,000 cycles, when usage is limited to 80 % of its initial capacity. Over the lifespan of 6,000 cycles, the storage capacity will degrade to 80 % of the original capacity. Eq. (11) says the storage system capacity ($C_s$) is equal to the storage system size ($E$) divided by the depth of discharge ($DoD$).

$$C_s = \frac{E}{DoD} \quad (11)$$

### GRID ELECTRICITY IMPORT

When both generation and storage cannot meet the demand, grid electricity is imported to meet the demand. Eq. (12) says the grid electricity import ($E(t)$) is equal to the unmet demand energy (($D(t) - G(t))\Delta t$) minus the remaining energy that can be extracted from storage ($S(t)/\eta_d$). Grid electricity is only imported when there is excess demand ($G(t) - D(t) < 0$) and storage is about to empty ($S(t+ \Delta t) \leq 0$). Otherwise, the system has enough stored energy or surplus generation to cover the demand, and electricity import is zero.

### COST AND PROJECTION

The cost of the hybrid renewable system is evaluated using LCOE (Levelized Cost of Electricity). Eq. (13) says LCOE is equal to the sum of annualized solar PV system cost ($P_{PV}$), lithium storage system cost ($P_S$), and electricity import cost ($P_E$), divided by the annual demand ($D$). Annualized solar PV system cost is equal to solar PV system capacity ($C_{PV}$) multiplied by the annualized solar PV system installed cost ($I_{PV}$) and O&M (Operation and Maintenance) cost ($O_{PV}$). The installed cost is annualized via the capital recovery factor equation, consisting of the discount rate ($r$) and solar PV system's lifespan ($n_{PV}$). The annualized lithium storage system cost is calculated in the same fashion, where $C_s$, $I_s$, $O_s$, and $n_s$ are lithium storage system's capacity, installed cost, O&M cost, and lifespan, respectively. The annual cost of electricity import ($C_E$) is equal to electricity import ($E(t)$) multiplied by the wholesale price ($P_E(t)$), summed over the year. The annual demand ($D$) is equal to hourly demand ($D(t)$) multiplied by an hour ($\Delta t$), summed over the year.

$$LCOE = \frac{P_{PV} + P_S + P_E}{D} \quad (13)$$

$$Where: P_{PV} = C_{PV} \left( I_{PV} \frac{r(1+r)^{n_{PV}}}{(1+r)^{n_{PV}} + 1} + O_{PV} \right)$$

$$P_S = C_S \left( I_S \frac{r(1+r)^{n_S}}{(1+r)^{n_S} + 1} + O_S \right)$$

$$P_E = \sum_{t=0}^{T} E(t)P_E(t)$$

$$D = \sum_{t=0}^{T} D(t)\Delta t$$

The lifespan of a solar PV system is 30 years, while a lithium storage system is 15 years (NREL, 2021). The discount rate is 3 %, consisting of 2.5 % annual inflation and 0.5 % base interest rate, both obtained from the UK average between 2011 and 2019 (World Bank, 2021; Bank of England, 2021). In 2019, solar and

$$E(t) = \begin{cases} (D(t) - G(t))\Delta t - \frac{S(t)}{\eta_d} , When \ S(t + \Delta t) \leq 0 , and \ G(t) - D(t) < 0 \\ 0, \quad Otherwise \end{cases} \quad (12)$$

storage's installed and O&M costs ($I_{PV}$, $O_{PV}$, $I_S$, $O_S$) are assumed to be EUR892/kW, EUR8.76/kW, EUR388/kWh, and EUR9.7/kWh, respectively (IRENA, 2020; NREL, 2021). These costs are projected to 2050 by NREL via an aggregation of multiple published projections (NREL, 2021). The projection predicts by 2030, solar's installed and O&M costs will decrease to 56 % and 73 % of 2019 level, respectively, and storage costs will decrease to 42 %. By 2050, solar's installed and O&M costs will lower to 46 % and 65 %, respectively, and storage costs will lower to 32 %. The analysis needs costs projected to 2100, so the cost reduction rate between 2050 and 2100 is assumed to be the same as that between 2030 and 2050. By 2100, solar's installed and O&M costs will reduce to 21 % and 48 %, respectively, and storage costs will reduce to 5 %. The cost projections are presented in Figure 4.

Great Britain's hourly wholesale electricity price ($P_E(t)$) are sourced from the Open Power System Data platform (OPSD, 2021). The price at each hour is calculated from a five-year average (2015 to 2019) to reduce anomalies. Figure 5 shows the hourly electricity price profile. Electricity price is generally higher during winter, and the price variation is also greater. The average electricity wholesale price over the year is EUR53/MWh.

**SIMULATION**

The analysis uses a hybridized analytical and iterative method to find the optimal solar PV and lithium storage system sizes. Figure 6 details the steps of the iterative sizing method. First, the method imports hourly electricity demand and solar capacity factor profiles shown in Figure 2 and 3. Then, the maximum solar PV system capacity is calculated using Eq. (4), and the solar size search range is set between zero and the maximum. The method iterates through the search range at an increment of 10 MW. At each solar size iteration, the solar generation profile is obtained via Eq. (3). Then, the storage profile is established using Eq. (5), and critical points in the profile are identified using Eq. (6) to form the difference matrix in Eq. (7). The maximum lithium storage system size is calculated via Eq. (8) using the difference matrix. The storage size search range is set between zero and the maximum, and the method iterates at a 10 MWh increment through the search range.

Within each storage size iteration, the method first iterates through all hours of the year, calculating the real storage level at each hour using Eq. (3) while constrained by Eq. (9). The real storage profile's ending storage level is the sustainable starting storage level, as stated by Eq. (10). Then, the method iterates through each hour of the year again, and simulates a new real storage profile that starts at the sustainable starting level. The grid electricity import is calculated from the real storage profile using Eq. (12). After the time iteration, storage size is converted into storage system capacity via Eq. (11), and the simulation yields a combination of solar PV system capacity, lithium storage system capacity, and grid electricity import. The combination's LCOE is calculated using Eq. (13). After the storage and solar size iteration, the LCOE of all combinations are calculated, and the lowest LCOE combination is selected as optimal. This process is repeated with solar and storage cost projections shown in Figure 4 to obtain the optimal combination for each year between 2019 and 2100.

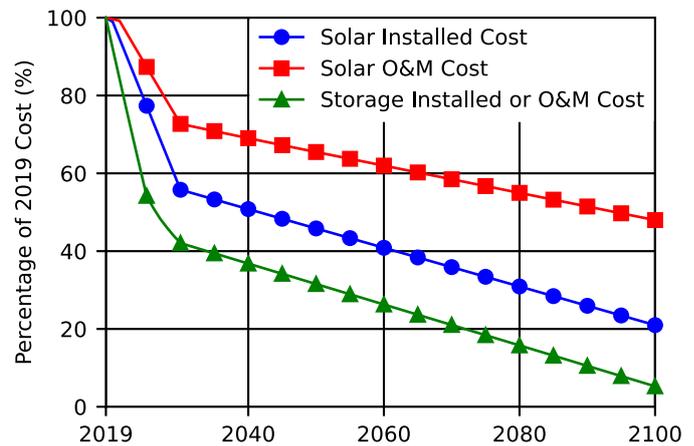

Figure 4. Projection of solar and storage's installed and O&M costs (NREL, 2021).

## Result and Discussion

The iterative method yields solar PV system capacity, storage system capacity, and grid electricity import combinations that can meet the demand. These combinations form a three-dimensional design space, shown in Figure 7. At the top of the figure, the data point "Only Grid" shows with no solar or storage, demand needs to rely solely on the grid, and thus grid electricity import is the highest. At the back-right of the figure, the data curve "Only Solar and Grid" shows that without storage, increasing solar system capacity can reduce grid electricity requirement to a certain extent. After which, increasing solar no longer reduces grid electricity import. This is because solar has no generation at night, and without storage, the nightly demand needs to rely on the grid.

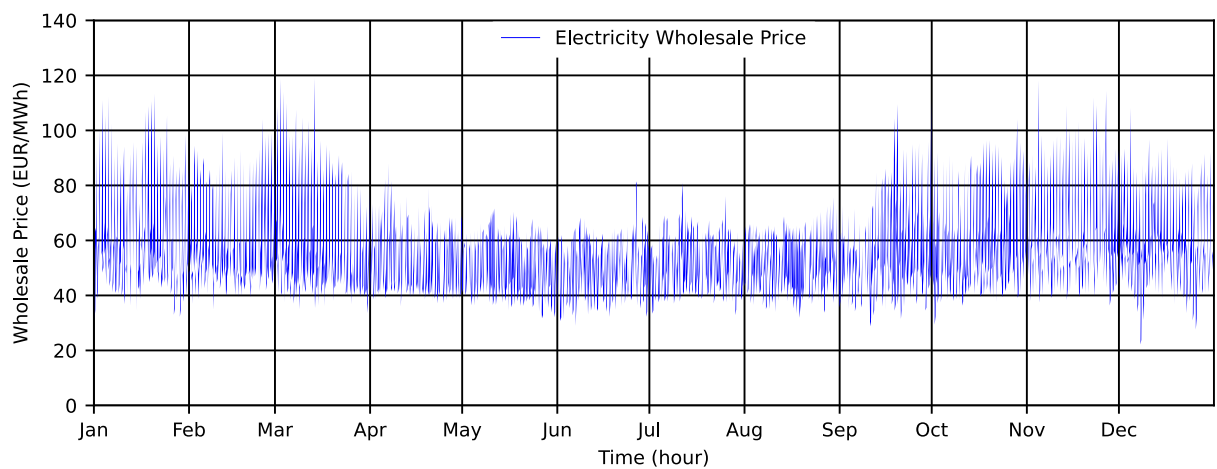

Figure 5. Great Britain's hourly wholesale electricity price in a year (OPSD, 2021).

At the back-left of the figure, the data curve "Max Storage with Solar and Grid" shows maximum storage capacity increases as solar capacity increases. This is because a larger solar capacity creates more surplus generation that can be stored, thus, increasing the maximum storage capacity. At the left of the figure, data point "Max Storage with Solar" shows the first solar and storage capacity that can meet the demand without grid electricity. The total solar generation is just enough to meet the total demand. With no over-generation, less demand is directly met by solar, and more demand needs storage, causing the large storage capacity requirement.

At the bottom of the figure, the data curve "Only Solar and Storage" shows the solar and storage capacities that can cover demand without grid electricity import. As solar capacity increases, the storage capacity requirement decreases. This is because more demands are directly met via solar generations, and less are met through storage, decreasing the storage requirement. However, the storage requirement reduction has a limit. Solar does not generate energy at night, and without grid electricity, storage is needed to meet the nightly demand.

In the middle of the figure, the data curve "Transition" defines a transition region. In the region right of the transition, increasing storage capacity can greatly reduce grid electricity import. This is because each storage capacity is utilized for most days during the year; they are charged during the day and discharged at night. The high utilization provides more energy to the demand, and less demand needs grid electricity. After the transition, increasing storage capacity reduces electricity import at a much slower pace. This is because the additional capacity is only utilized once a year; they are charged during the summer and discharged during fall and winter. The lower utilization provides less energy for demand, resulting in less reduction on electricity import.

Based on the design space, each combination's LCOE is calculated, and the lowest LCOE combination is selected as the optimal. This process is repeated for each year between 2019 and 2100, and the results are shown in Figure 8. The optimal system in 2019 consists of 140 MW of solar generation with no storage, and the grid supplies 78 % of the energy. The result shows while solar PV has become economically competitive, lithium storage remains too expensive. As time passes, solar and storage system prices will continue to drop, and they will take up greater portions of the hybrid renewable system. In 2050, the optimal solar capacity doubles to 270 MW, and grid electricity import reduces to 64 %.

Lithium battery storage becomes part of the optimal system in 2074. By then, projected battery system costs have reduced to 19 % of the 2019 level, and the solar PV system's installed and O&M costs have reduced to 34 % and 57 %, respectively. The optimal system consists of 330 MW of solar PV and 90 MWh of lithium storage, while grid electricity import is reduced to 59 %. After storage becomes part of the optimal system, optimal solar and storage capacity begin to increase rapidly. By 2100, the optimal system consists of 620 MW of solar and 990 MWh of storage, while electricity import is reduced to 33 %. Between 2019 and 2100, the optimal system's grid electricity reliance is reduced by 45 %, and the LCOE is reduced from EUR54/MWh to EUR34/MWh.

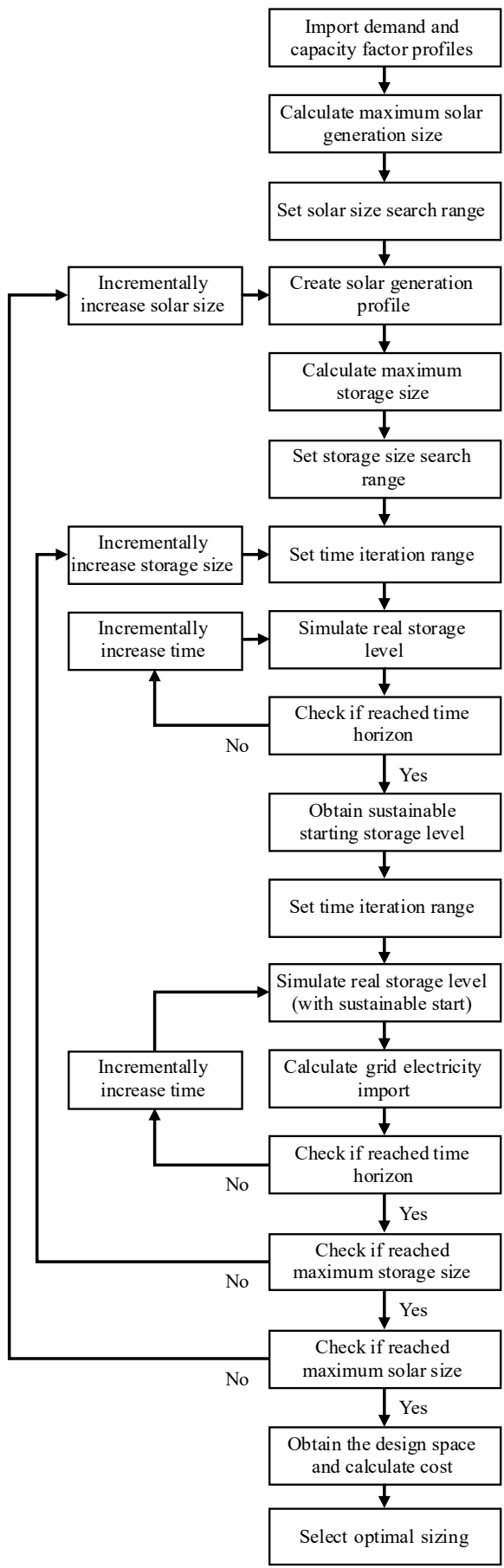

*Figure 6. Flowchart of the hybridized analytical and iterative method for sizing hybrid renewables.*

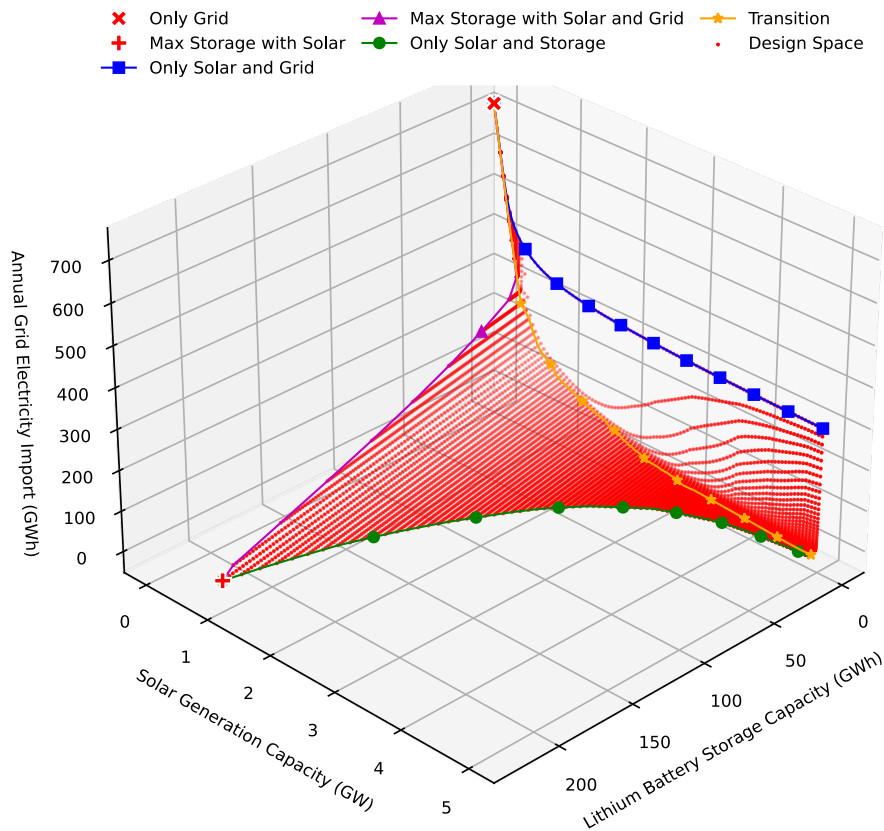

*Figure 7. Design space with solar capacity, storage capacity, and annual grid electricity import for Oxford.*

Solar and storage will take up greater portions of the hybrid renewable system as their costs come down. However, the solar and storage capacity increase will eventually slow down, as some demands are too expensive to meet by solar and storage alone. For example, nightly winter demand either requires large storage to store surplus generation from the summer, or large solar to generate enough energy during the winter. Thus, there will always be a need for grid electricity, which provides flexibility at a competitive price.

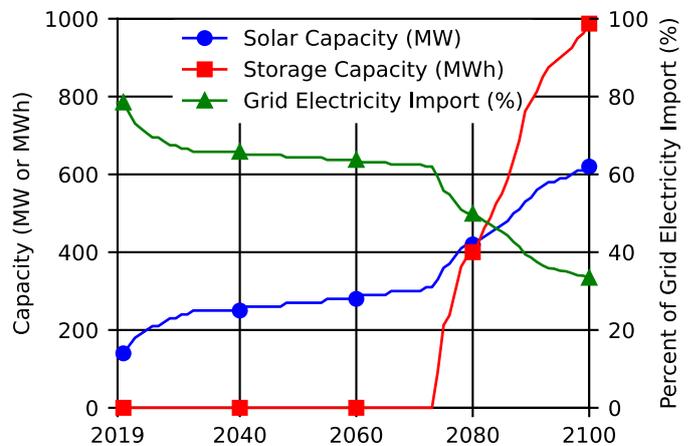

*Figure 8. Optimal solar capacity, storage capacity, and grid electricity import for Oxford.*

## Conclusion

This paper presents a hybridized analytical and iterative method to size solar PV and lithium storage systems, reducing grid reliance in buildings. The mathematical model for each component is developed, and the hybridized method is used for sizing. The method iterates through the solar size search range. At each solar size, the method iterates through the storage size search range, and simulates the hybrid renewable system. The method outputs combinations of solar PV system capacity, lithium storage system capacity, and grid electricity import that can meet the demand. Each combination's LCOE is calculated, and the lowest LCOE combination is the optimal sizing. This process is repeated for each year between 2019 to 2100 using cost projections.

The method is applied to Oxford city, and the result shows the 2019 optimal system sources 22 % of its energy from solar PV (140 MW) and the rest from the grid. Lithium battery storage becomes viable in 2074, and the optimal system sources 41 % of its energy from lithium storage (90 MWh) and solar (330 MW). In 2100, the optimal system sources 67 % of its energy from solar (620 MW) and storage (990 MWh). Between 2019 and 2100, the LCOE reduced from EUR54/MWh to EUR34/MWh, while grid reliance decreased by 45 %.

Currently, solar PV is economically competitive, but lithium storage cost is still too high. As solar and storage prices continue to drop, they will become more viable and take up greater portions of the energy system. However, there will always be a need for grid electricity import. The grid provides flexibility and can meet some demands that are too costly for solar and storage. Thus, solar and storage will become a greater part of the future energy system, but there will always be a need for the grid.